\documentclass{llncs}

\usepackage{hyperref}
\usepackage{eso-pic}
\AddToShipoutPicture*{
\put(133,110){
\scriptsize
Proceedings of \href{http://arxiv.org/abs/1406.1510}{11\textsuperscript{th} Workshop on Constraint Handling Rules (CHR 2014)}, 
paper CHR/2014/2
}}

\usepackage{graphicx}
\usepackage{longtable}

\begin{document}

\title{From XML Schema to JSON Schema:\\Translation with CHR}
\author{Falco Nogatz and Thom Fr{\"u}hwirth}
\institute{Faculty of Engineering and Computer Sciences, Ulm University, Germany\\
	\email{\{falco.nogatz,thom.fruehwirth\}@uni-ulm.de}}

\maketitle

  \begin{abstract}
Despite its rising popularity as data format especially for web
services, the software ecosystem around the JavaScript Object Notation (JSON)
is not as widely distributed as that of XML. For both data formats there exist
schema languages to specify the structure of instance
documents, but there is currently no opportunity to translate already existing XML
Schema documents into equivalent JSON Schemas.

In this paper we introduce an implementation of a language translator. It takes
an XML Schema and creates its equivalent JSON Schema document. Our approach
is based on Prolog and CHR. By unfolding the 
XML Schema document into CHR constraints, it is possible to specify the
concrete translation rules in a declarative way.
  \end{abstract}



\section{Introduction}

XML, the Extensible Markup Language \cite{bray1997extensible}, is today one
of the most used formats to save and exchange structured data. Being a recommendation
of the World Wide Web Consortium (W3C) since 1998, 
a large software ecosystem has been evolved, including
data formats to specify the schema of XML documents. One of them is the XML Schema
Definition (XSD) \cite{part2001structures}.

Since its proposal in 2006, there is an alternative data format especially used in
web services: JSON, the JavaScript Object Notation. Its formal language to specify the
format of a JSON document, called JSON Schema, is still in draft status \cite{jsondraft04}.
Although there are validation tools implementing the IETF draft, the
number of JSON Schemas used in practice is still moderate. One of the reasons is
that there is currently no mechanism to translate an already existing XML Schema
into equivalent JSON Schema.

As an application of XML, XSD documents are valid XML instances.
Although JSON Schema is JSON-based as well, the naive approach of using an
already existing XML to JSON translator as published by \cite{nathan2007convert} would not
result in a valid JSON Schema document. To satisfy the
Core Meta-Schema \cite{jsonmetaschema}, the demanded translator has to
provide some additional logic, extending the general problems of translating
XML to JSON instances as presented in \cite{jxon}.

In this paper, we propose an approach for an XSD to JSON Schema language
translator based on Prolog and Constraint Handling Rules (CHR) \cite{CHRfruehwirth}.
The translator unfolds
a given XML Schema into CHR constraints. By creating a CHR constraint for every
XSD node it is possible to specify the concrete translation rules of
common XML Schema fragments in a declarative way in form of CHR rules.

The paper is organized as follows. In Section \ref{Preliminaries},
we will give an example to illustrate the
problem and we will determine the considered versions of the XSD and JSON Schema
specifications. The paper continues by presenting the introduced CHR constraints.
In Section \ref{Translation_Process} the overall translation process is presented.
Finally, the paper ends with concluding remarks in Section \ref{Conclusion}.

\section{Preliminaries}
\label{Preliminaries}

The aim of this work is to create a Prolog/CHR module that offers a
predicate \texttt{xsd2json(XSD,JSON)} which holds the equivalent
JSON Schema as \texttt{JSON} for a given \texttt{XSD} instance. Before getting
into the concrete translation process we want to introduce the used techniques
and specify the scope of this tool. In what follows we explain the
problem instance by giving an example of a simple XSD and its
expected translated JSON Schema equivalent.

\subsection{Problem Definition}

Following the formal description of the XML Schema language \cite{brown2001xml},
an XML Schema consists of four components: elements (\texttt{xs:element} nodes),
simple types (\texttt{xs:simpleType} nodes), 
complex types (\texttt{xs:complexType} nodes) and 
attributes (\texttt{xs:attribute} nodes). Because the also introduced attribute groups and
model groups are only placeholders in complex type definitions, we will omit those components
for our translator. In Section \ref{Fragment_Translation}, we will 
introduce translation rules for each of the four given components,
depending on their structure and values.

Although the XML Schema 1.1 Specification has been the official W3C recommendation since April
2012, we restrict ourselves to the XML Schema 1.0 Specification. The
more up-to-date specification primarily introduces conditional types and assertions
based on XPath expressions. Since there is currently no XPath equivalent for JSON, it would
not be possible to translate those new XPath-based elements at all.

For the target language JSON Schema we refer to the latest version of the specification,
Draft 04 \cite{jsondraft04}, which is 
already supported by a number of JSON validators in multiple languages.
A list of current implementations can be found in \cite{JSONschemaSoftware}.

\subsection{Problem Instance Example}
\label{Problem_Example}

As a motivating example, we will consider a small XML document, as shown in
Figure \ref{XML_example}, and its related XSD, as specified in Figure
\ref{XSD_example}.

\begin{figure}
\begin{minipage}[t]{.5\textwidth}
\begin{verbatim}
<?xml version="1.0" ?>
<percentages>
  <value>99</value>
  <value>42</value>
  <value>0</value>
</percentages>
\end{verbatim}
\vspace*{-0.5cm}
\caption{Example XML}
\label{XML_example}
\end{minipage}
\hfill
\begin{minipage}[t]{.5\textwidth}
\begin{verbatim}
{
  "value": [ 99, 42, 0 ]
}
\end{verbatim}
\vspace*{-0.5cm}
\caption{JSON document, valid against the JSON Schema of Figure \ref{JSV_example}}
\label{JSON_example}
\end{minipage}
\end{figure}

The aim of the language translator is to create an equivalent JSON Schema
of the XSD given in Figure \ref{XSD_example}. It should respect the following the semantics:
\begin{itemize}
\item There is a list of values.
\item The list contains at most five values.
\item Every value must be a nonnegative integer.
\end{itemize}

Following the XSD specification in \cite{brown2001xml}, there is additional
information implicitly given: By omitting the \texttt{minOccurs} attribute in
an \texttt{xs:element} within an \texttt{xs:sequence} its default value \texttt{1}
is used, so the list has to contain at least one value. 

The equivalent JSON Schema that ensures these constraints is shown in 
Figure \ref{JSV_example} and its corresponding JSON document in Figure \ref{JSON_example}.
The \texttt{percentages} node of the XML document has no equivalent
in the JSON Schema instance. This is caused by the circumstance that the \texttt{percentages} element
adds no constraints and therefore might only be used to create a valid XML document, which
requires a single root element. The language translator uses such assumptions to create a
simple, but appropriate JSON Schema.

\subsection{CHR Constraints}
\label{CHR_Constraints}

To provide translation rules for concrete XSD fragments, we use a combination
of the logic programming languages Prolog and
CHR \cite{CHRfruehwirth}\cite{Fruehwirth:2011:CHR:2011970}. This
enables us to specify the translation rules in a declarative way. Since for each
XSD node a new CHR constraint will be generated, it is possible to create
CHR rules referencing constraints by their characteristics without having to 
implement the tree traversal of the XSD document.

We use CHR with Prolog as its host language. The suggested implementation can be found
online at \url{https://github.com/fnogatz/xsd2json} and 
has been tested with the CHR library for SWI-Prolog \cite{fruehwirth2012swi}. To hold 
the information of a given XSD term 
we introduce the following CHR constraints:

\begin{itemize}
\item \texttt{node(Namespace,Name,ID,Children\_IDs,Parent\_ID)}\\
For each XML node in the XSD document a new \texttt{node/5} constraint is generated,
holding its namespace and tag name. To obtain a reference, a unique identifier is added as
well as the list of its parent's and children's identifiers.

\item \texttt{node\_attribute(ID,Key,Value,Source)}\\
For each XSD attribute a new \texttt{node\_attribute/4} constraint is propagated,
holding its name as \texttt{Key}, its \texttt{Value} and 
the identifier of the related \texttt{node/5} constraint. The \texttt{Source} is
\texttt{source} for explicitly set and \texttt{default} for inherited attributes.
For example \texttt{maxOccurs="5"} of the innermost \texttt{xs:element}
of Figure \ref{XSD_example} is mapped to
a constraint \texttt{node\_attribute(\_ID,maxOccurs,5,source)}.

\item \texttt{text\_node(ID,Text,Parent\_ID)}\\
If an element's child is simply a text and no nested XML node, a \texttt{text\_node/3}
constraint is generated. It gets a unique identifier like a regular child node and
holds the text as well as the identifier of its parent element.
\end{itemize}

All translated fragments are stored in \texttt{json(ID,JSON)} constraints, holding
the \texttt{JSON} Schema of the XSD node with the identifier \texttt{ID}. Because
the entire JSON Schema is built step by step, the innermost fragments of the XSD
propagate the first \texttt{json/2} constraints. These will be picked up for
the translation of their parent elements, resulting in a JSON Schema for the
entire XSD.

\begin{figure}
\begingroup
    \fontsize{8pt}{9pt}\selectfont
\begin{minipage}[t]{.6\textwidth}
\begin{verbatim}
<?xml version="1.0" ?>
<xs:schema
  xmlns:xs="http://www.w3.org/2001/XMLSchema">
  <xs:element name="percentages">
    <xs:complexType>
      <xs:sequence>
        <xs:element
          name="value"
          maxOccurs="5"
          type="xs:nonNegativeInteger" />
      </xs:sequence>
    </xs:complexType>
  </xs:element>
</xs:schema>
\end{verbatim}
\vspace*{-0.5cm}
\caption{Possible XSD for XML of Figure \ref{XML_example}}
\label{XSD_example}
\end{minipage}
\hfill
\begin{minipage}[t]{.4\textwidth}
\begin{verbatim}
{
  "type": "object",
  "properties": {
    "value": {
      "type": "array",
      "items": {
        "type": "integer",
        "minimum": 0, 
        "exclusiveMinimum": false
      },
      "minItems": 1,
      "maxItems": 5
    }
  },
  "required": [ "value" ]
}
\end{verbatim}
\vspace*{-0.5cm}
\caption{Tanslated JSON Schema, based on the XSD of Figure \ref{XSD_example}}
\label{JSV_example}
\end{minipage}
\endgroup
\end{figure}
\vspace*{-1cm}

\section{Translation Process}
\label{Translation_Process}

\begin{figure}[htp]
\begin{center}
  \includegraphics[width=\linewidth]{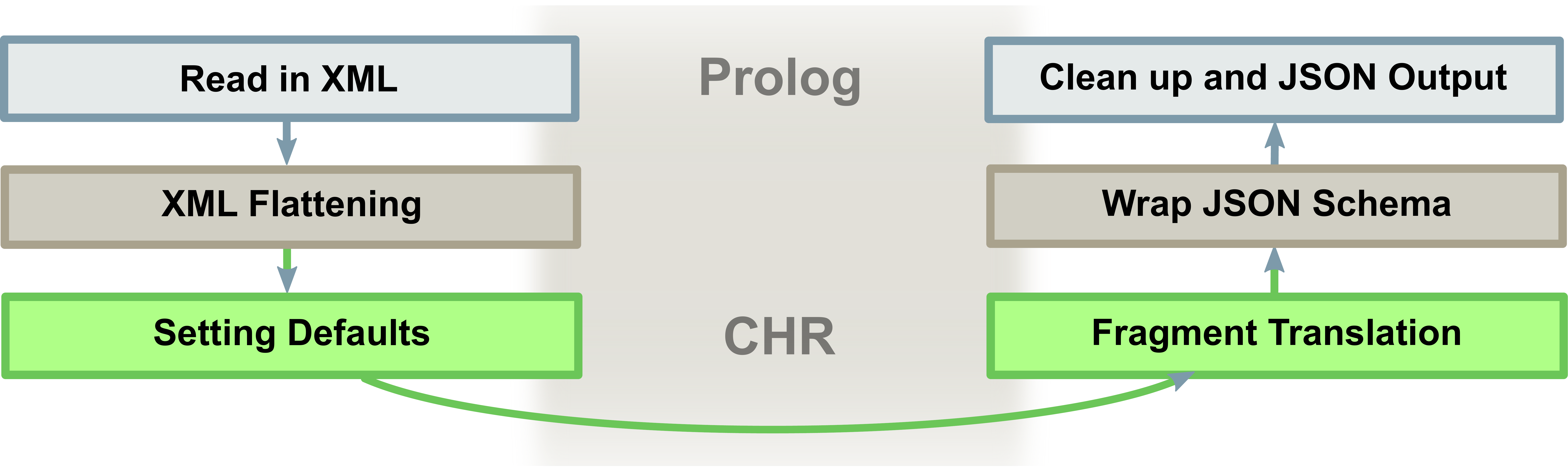}
  \caption{Steps of the overall translation process}
  \label{fig:translation:steps}
\end{center}
\end{figure}

The overall translation process can be split into six subtasks as illustrated
in Figure \ref{fig:translation:steps}. The different steps can be distinguished by
their function as well as by the used programming language.

In the following we will present the various steps. The main part of the translator, the
translation rules of XSD fragments, is introduced in Section \ref{Fragment_Translation}.

\subsection{Read in XML Schema into Prolog}

SWI-Prolog provides a wide support for working with XML documents.
By use of its SGML/XML parser \cite{SWIprologSGML}, an XSD document can be read in as
a nested Prolog term. Figure \ref{lst:xsdProlog} shows the term generated
by the built-in \linebreak\texttt{load\_structure/3} predicate \cite{fruehwirth2012swi} for the XSD
of Figure \ref{XSD_example}.

\begin{figure}
\begin{verbatim}
[ element(
    'http://www.w3.org/2001/XMLSchema':schema,        % namespace and name
    [ xmlns:xs='http://www.w3.org/2001/XMLSchema' ],  % attributes
    [ element(                                        % nested elements
        'http://www.w3.org/2001/XMLSchema':element,
        [ name=percentages ],                         % attributes
        [ ... ])                               % the other nested elements
    ])]
\end{verbatim}
\vspace*{-0.5cm}
\caption{Nested Prolog term of the XSD document of Figure \ref{XSD_example}}
\label{lst:xsdProlog}
\end{figure}

\subsection{XML Flattening}

This nested Prolog term can be traversed recursively to propagate the related \texttt{node/5}, \texttt{node\_attribute/4} and \texttt{text\_node/3}
constraints. Their positions are retained by their unique identifiers and references
to parent and child nodes.

\subsection{Setting Defaults}

Because Prolog's XML parser will only read in explicitly set attributes, we have to
add the default attributes as shown in Section \ref{Problem_Example}. The
translation rules used in the next step refer to attributes like \texttt{minOccurs}
and \texttt{maxOccurs}, which can be omitted. To ensure these optional attributes
are always present, we propagate a \texttt{node\_attribute/4} with
the \texttt{Source} set to \texttt{default}, as mentioned in
Section \ref{CHR_Constraints}. If there is an identical \texttt{node\_attribute/4}
constraint with its last component set to \texttt{source}, the default one
is removed by a CHR simpagation rule.

\subsection{Fragment Translation}
\label{Fragment_Translation}

Before examining the most important step, we will have to a look at
the intended result of the overall translation process: a Prolog representation of
JSON Schema. Like for XML, SWI-Prolog comes with
a library to serialize JSON. With the \texttt{http/json} library \cite{fruehwirth2012swi}
a JSON object is represented by \texttt{json(L)}, in which \texttt{L}
is a list of the form \texttt{[Key1=Value1,Key2=Value2,\dots]}. JSON
arrays are represented by Prolog lists.

Referring to the constraints propagated in the steps before, we can translate
XSD fragments to their equivalent JSON Schema. All translation rules follow the
same form: They propagate a single \texttt{json/2} constraint
that holds the JSON Schema of this XSD fragment; the guard ensures that
all \texttt{node/5} and \texttt{node\_attribute/4} constraints are of the
XML Schema namespace. The rule's head contains the following parts:

\begin{itemize}
\item The \texttt{node/5} constraint for which the \texttt{json/2} representation
is generated. In most cases its children's \texttt{node/5} constraints are
referenced by using the \texttt{Parent\_ID} component.

\item Some \texttt{node\_attribute/4} and \texttt{text\_node/3} constraints, depending
on the XSD fragment that has to be translated.

\item Some \texttt{json/2} constraints to merge already translated fragments. This way
it is possible to generate the JSON Schema translation of an XSD node by
combining the translations of its child nodes.
\end{itemize}

Hereby the propagation rule to generate the JSON Schema for the innermost \texttt{xs:element}
of Figure \ref{XSD_example} is:

\begin{verbatim}
node(NS,element,ID,_C,_Parent), node_attribute(ID,type,With_NS,_)
  ==> xsd_namespace(NS), valid_xsd_type(With_NS,Type)
  |   convert_xsd_type(Type,JSON), json(ID,JSON).
\end{verbatim}

This rule applies if the \texttt{xs:element} node has an XSD namespace and
is of a primitive XSD data type.
XML Schema provides various predefined data types. Although the number
of data types defined for JSON is limited, it is possible to restrict
them similarly to constraining facets \cite{xmldatatypes} in XSD.
Therefore we can define a \texttt{convert\_xsd\_type/2} predicate by providing JSON's equivalents
of all predefined XML data types like in Table \ref{tbl:rules:types} in excerpts.

\begin{longtable}{c|c}
\caption{Translation of simple XSD data types (extract of \cite{nogatz})}
\label{tbl:rules:types}\\

\textbf{XSD primitive type} & \textbf{JSON Schema type definition}\\
\hline
\begin{minipage}{.5\textwidth}
  \verb|xs:string|
\end{minipage} &
\begin{minipage}{.45\textwidth}
\verb| { "type": "string" }|
\end{minipage}\\

\hline
\begin{minipage}{.5\textwidth}
\verb|xs:float, xs:double, xs:decimal|
\end{minipage} &
\begin{minipage}{.45\textwidth}
\verb| { "type": "number" }|
\end{minipage}\\

\hline
\begin{minipage}{.5\textwidth}
\verb|xs:nonNegativeInteger|
\end{minipage} &
\begin{minipage}{.45\textwidth}
\begin{verbatim}
 { "type": "integer",
   "minimum": 0,
   "exclusiveMinimum": false }
\end{verbatim}
\end{minipage}
\end{longtable}

XSD's primitive data types can be restricted by constraining facets \cite{xmldatatypes},
for example to specify all possible values for a string.
These facets can be translated by using a similar
table of equivalents, which are collected in \cite{nogatz}.

The primitive data types and constraining facets apply only to \texttt{xs:attribute}
and certain \texttt{xs:element} nodes. However, an XSD is 
more than the definition of simple types: via \texttt{xs:complexType}
nodes attributes of elements can be specified as well as their child nodes.
The occurence of \texttt{xs:element} within an \texttt{xs:sequence} as shown in
Figure \ref{XSD_example} is a common structure in XSD documents.
Therefore we have to translate nested XSD nodes as the last part of this step.
The general approach has already been introduced before: depending on specific
\texttt{node/5}, \texttt{node\_attribute/4} and \texttt{text\_node/3} constraints
and sometimes already translated fragments given as \texttt{json/2} constraints, 
we compose the translation of an XSD node. 

As an example we introduce the actual translation rule for the nested\linebreak
\texttt{xs:sequence}/\texttt{xs:element} structure of the example in 
Figure \ref{XSD_example}:

\begingroup
    \fontsize{8pt}{9pt}\selectfont
\begin{verbatim}
node(NS1,sequence,Sequence_ID,_SC,_SP), node(NS2,element,El_ID,_EC,Sequence_ID),
    json(El_ID,Element_JSON), node_attribute(El_ID,name,Element_Name,_)
    node_attribute(El_ID,minOccurs,MinOccurs,_O),      % _O to match both origins
    node_attribute(El_ID,maxOccurs,MaxOccurs,_O),      %   'default' and 'source'
  ==>
    xsd_namespace(NS1), xsd_namespace(NS2)             % valid XSD namespaces?
  | JSON = [ type=object,                              % build JSON object
             properties=json([
               Element_Name=json([ type=array,
                                   items=Element_JSON,
                                   minItems=MinOccurs,
                                   maxItems=MaxOccurs
               ]) ]) ],
    (MinOccurs_Number >= 1, Full_JSON = [required=[Element_Name]|JSON];
      MinOccurs_Number < 1, Full_JSON = JSON),         % required property?
    json(Sequence_ID,json(Full_JSON)).                 % propagate JSON
\end{verbatim}
\endgroup

Many applications of nested structures have been identified, documented in \cite{nogatz}
and implemented by similar propagation rules. Because a 
\texttt{node/5} constraint might apply to multiple CHR rules, there can be
various \texttt{json/2} constraints with the same identifier. These are merged
by a simpagation rule with the help of a self-defined \texttt{merge\_json}
Prolog predicate.

\subsection{Wrap JSON Schema}

The previous step terminates as soon as the root element of the given XSD has
been translated and its \texttt{json/2} constraints have been merged. In addition
the globally defined type definitions are merged into the \texttt{definitions}
object of the root's \texttt{json/2} constraint.

\subsection{Clean-up and JSON Output}

Finally, the created JSON Schema object is cleaned up: 
in the creation process, the names of XML attributes (specified as
\texttt{xs:attribute} in the XSD) were prefixed with an \texttt{@} symbol. If there
is no \texttt{xs:element} in this \texttt{xs:complexType} with the same name, 
the attribute's \texttt{@}-prefix is removed.

\section{Conclusion}
\label{Conclusion}

In this work, a language translator to convert XML Schema to an equivalent JSON Schema was
implemented. The entire implementation is available online at 
\url{https://github.com/fnogatz/xsd2json}, its detailed concept can be found in \cite{nogatz}.
As the \texttt{xsd2json} Prolog/CHR module was developed in
a bottom-up approach, it also provides a test framework and a large number of
test cases.

The \texttt{xsd2json} module is still under development to be applicable for
all XML Schema instances. Due to the lack of various features in JSON Schema it might
not be possible to support all constraining semantics. For example, there is currently
no XPath-like way to address a specific property inside a nested JSON document.
Therefore, the XSD elements \texttt{xs:key}, \texttt{xs:keyref} and \texttt{xs:unique}
cannot be supported as well as the new features like \texttt{xs:assertion} 
introduced by the most up-to-date XML Schema 1.1 Specification.

Another missing feature is the handling of referenced XSD documents. While the
current implementation respects multiple namespaces, it translates only
a single file. Therefore \texttt{xs:import} and \texttt{xs:include} are not supported.

\bibliographystyle{splncs}

\end{document}